# Studies of LL-type 500MHz 5-cell superconducting cavity at SINAP*

HOU Hong-Tao(侯洪涛)[1,3], MA Zhen-Yu(马震宇)[1,3], MAO Dong-Qing(毛冬青)[1,3], FENG Zi-Qiang(封自强)[1,3], LUO Chen(罗琛)[1,3], SHI Jing(是晶)[1,3], WANG Yan(王岩)[1,3], LI Zheng(李正)[1,3], XU Kai(徐凯)[1,3], ZHAO Yu-Bin(赵玉彬)[1,3], ZHENG Xiang(郑湘)[1,3], ZHAO Shen-Jie(赵申杰)[1,3], ZHANG Zhi-Gang(张志刚)[1,2,3] , LIU Jian-Fei(刘建飞)[1,3;1)]
1 (Shanghai Institute of Applied Physics, Chinese Academy of Sciences, Shanghai 201800, China)
2 (University of Chinese Academy of Sciences, Beijing 100049, China)
3（Shanghai Key Laboratory of Cryogenics & Superconducting RF Technology, Shanghai 201800, China）

A low loss (LL) type 500 MHz 5-cell superconducting niobium prototype cavity with large beam aperture has been developed successfully including the optimization, the deep drawing and electron beam welding, the surface treatment and the vertical testing. The performance of the fundamental mode was optimized and the higher order modes were damped by adopting an enlarged beam pipe for propagation. Surface preparation or treatment including mechanical polishing, buffered chemical polishing and high pressure rinsing with ultra-pure water and so on was carried out carefully to promise a perfect inner surface condition. The vertical testing results show that the accelerating voltage higher than 7.5 MV was obtained while the quality factor was better than $1\times10^9$ at 4.2 K. No obvious multipacting or field emission was found during the test. However, a quench happened while increasing the field a little higher than 7.5 MV that at present limited the cavity performance.

## 1 Introduction

As the key component, many different kinds of multi-cell superconducting cavities have been designed or fabricated [1-9] aiming to the future high current accelerators such as Energy Recovery Linacs (ERL) and Free Electron Laser sources (FEL) or ERL based high power terahertz sources [10] and high average beam current synchrotron light sources [11]. ERL was proposed to be an alternative synchrotron radiation source [11-13] that might meet the users' desired requirement of extremely low emittance, super-brilliance and ultra-short bunches for smaller samples and shorter time window examination. The superconducting cavities have been proven to be the best solution for compact and high power high current accelerators.

A multi-cell superconducting cavity has the advantage to maximize the active acceleration length thus promises a small number of cryomodules needed within a shorter length, which can save the valuable straight sections for insertion devices. Because of their large aperture, superconducting cavities resonant at a low frequency adopted in the high current accelerators can suppress the wake field effects and higher order mode losses. The low frequency superconducting cavities have low surface resistance in BCS theory [14] compared to those cavities resonant at higher frequency such as 1.3 GHz. The challenge from a big cryogenic power loss for a continuous wave high current accelerator might be solved by adopting low frequency cavities. However, it will also bring another challenge from fabrication and surface treatment to make a perfect inner surface because of the large size of low frequency superconducting cavities.

The 500 MHz single cell niobium cavity has been developed in 2010 at Shanghai Institute of Applied Physics (SINAP) and the vertical test results show a good performance to reach the world level [15]. This kind of cavity can be used as injector rf cavities in an ERL facility which promotes the development of a 500 MHz multi-cell superconducting cavity. This paper firstly illustrates the rf design and the optimized parameters of the final cavity shape. The fabrication of niobium together with surface treatment applied and the pre-tuning of the niobium cavity are also described. Finally, the vertical testing results are presented.

## 2 Cavity design

The design of the 500 MHz 5-cell cavity follows some rules (i) a large beam iris to reduce wake filed effects and increase the cell-to-cell coupling which is good for higher order modes coupling and low sensitive factor to mechanical error, (ii) minimization of peak surface electric and magnetic fields while keeping a large (r/Q)*G value of the fundamental mode to decrease the dissipated power, (iii) an enlarged beam pipe for higher order modes (HOMs) propagation and damping. The cavity shape was optimized to be low loss type having an elliptical contour to lower the sensitivity for multipacting and a straight line at the equator which is different to the original design [8]. The geometry parameters of inner half cell of the cavity are shown in figure 1, where $l$ is the cell length, $R_e$ is the equator radius, $R_i$ is the iris radius, $b_1/a_1$ is the aspect ratio of the equator ellipse, $b_2/a_2$ is the aspect ratio of the iris ellipse, $d$ is length of the straight section at the equator and α is the slope angle. The straight line at the equator helps the frequency control and balance the field flatness of the cavity. The cavity field flatness was reached better than 98% which was ensured by adjusting the straight line length to be 12.8 mm shorter at the equator of the end cells than that of the mid-cells. One significant benefit of

1) Corresponding author (email: liujianfei@sinap.ac.cn)
* Work is partially supported by National Natural Science Foundation of China (11175237)

this method is that it needs only one set of dies for both inner cell and end cell shaping and trimming. The length of the beam pipe 300 mm was determined to make the fundamental mode decay higher than 37 dB at the end of the beam pipe. The optimized values are listed in Table 1. A symmetry structure was adopted in the final cavity geometry. The cavity will be equipped with a coaxial type fundamental coupler which has been proven to be capable of handling a much higher input power and an adjustable coupling strength.

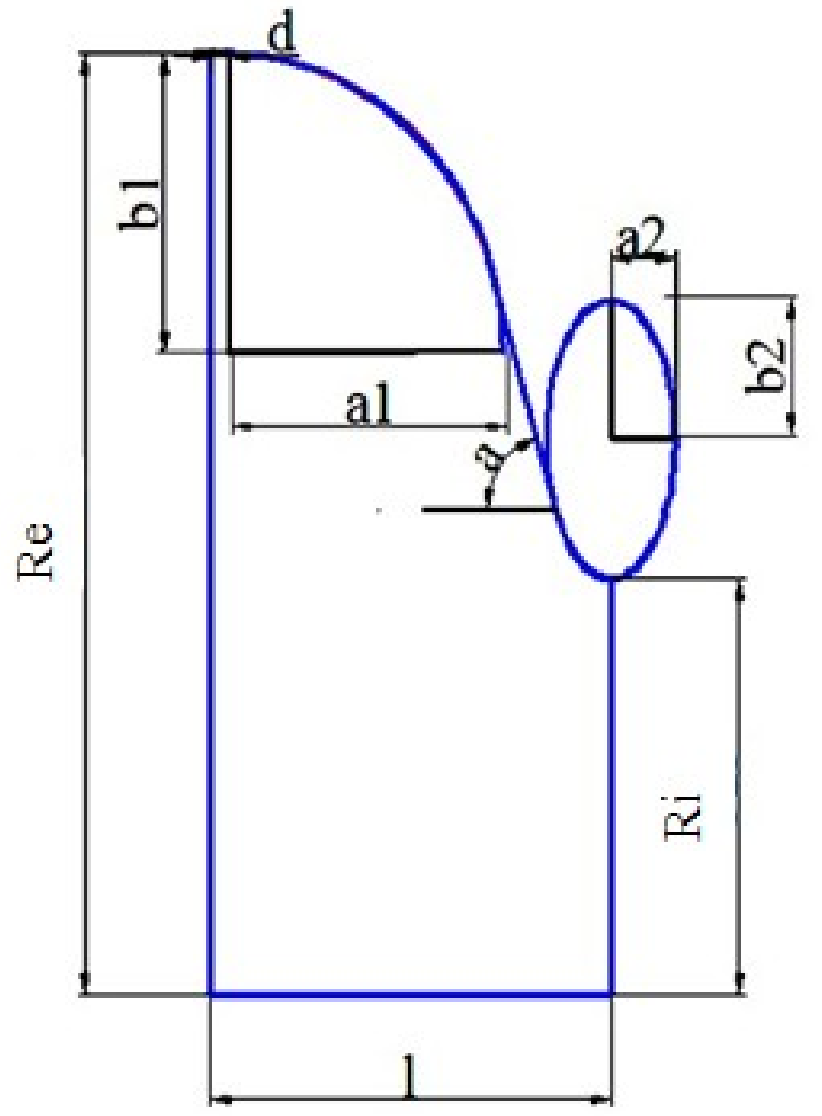

Figure 1 Geometry parameters of inner half cell

Table 1 Optimized parameters of the 500 MHz 5-cell cavity

| Parameters | Inner-cell | End-cell | Transition |
|---|---|---|---|
| $R_e$ / mm | 263.0 | 263.0 | 160.0 |
| $R_i$ / mm | 105.0 | 105.0 | 105.0 |
| $d$ / mm | 16 | 3.2 | / |
| $l$ / mm | 150 | 143.6 | 58.78 |
| $a_1$ / mm | 112.0 | 112.0 | 25.35 |
| $b_1$ / mm | 89.3 | 89.30 | 25.35 |
| $a_2$ /mm | 30.0 | 30.0 | 30.0 |
| $b_2$ / mm | 30.0 | 30.0 | 30.0 |
| $\alpha$ / ° | 90 | 90 | 70 |

Table 2 RF performance for the 500 MHz 5-cell cavity

| Parameters | Value |
|---|---|
| $f$ /MHz | 499.6 |
| $r/Q$ / Ω | 515.6 |
| $G$ / Ω | 275.5 |
| $k_{cc}$ / % | 3.18 |
| $E_p/E_{acc}$ | 2.51 |
| $H_p/E_{acc}$ / Oe/(MV/m) | 42.9 |
| Field Flatness / % | 98.9 |

With Superfish [16] and CST [17], the optimized rf performance of the 500 MHz 5-cell cavity is obtained and listed in Table 2. The ratio of peak surface to accelerating field $E_p/E_{acc}$ =2.51 and $H_p/E_{acc}$ =42.9 Oe/(MV/m). The cell-to-cell coupling is 3.18% which is good for propagating HOMs out the cells and a higher value to make the field distribution less sensitive to small dimensional variations to obtain a flat accelerating mode field. Figure 2 shows the 500 MHz 5-cell cavity fundamental field simulation results from Superfish and both field flatness calculation results by Superfish and CST are better than 98%.

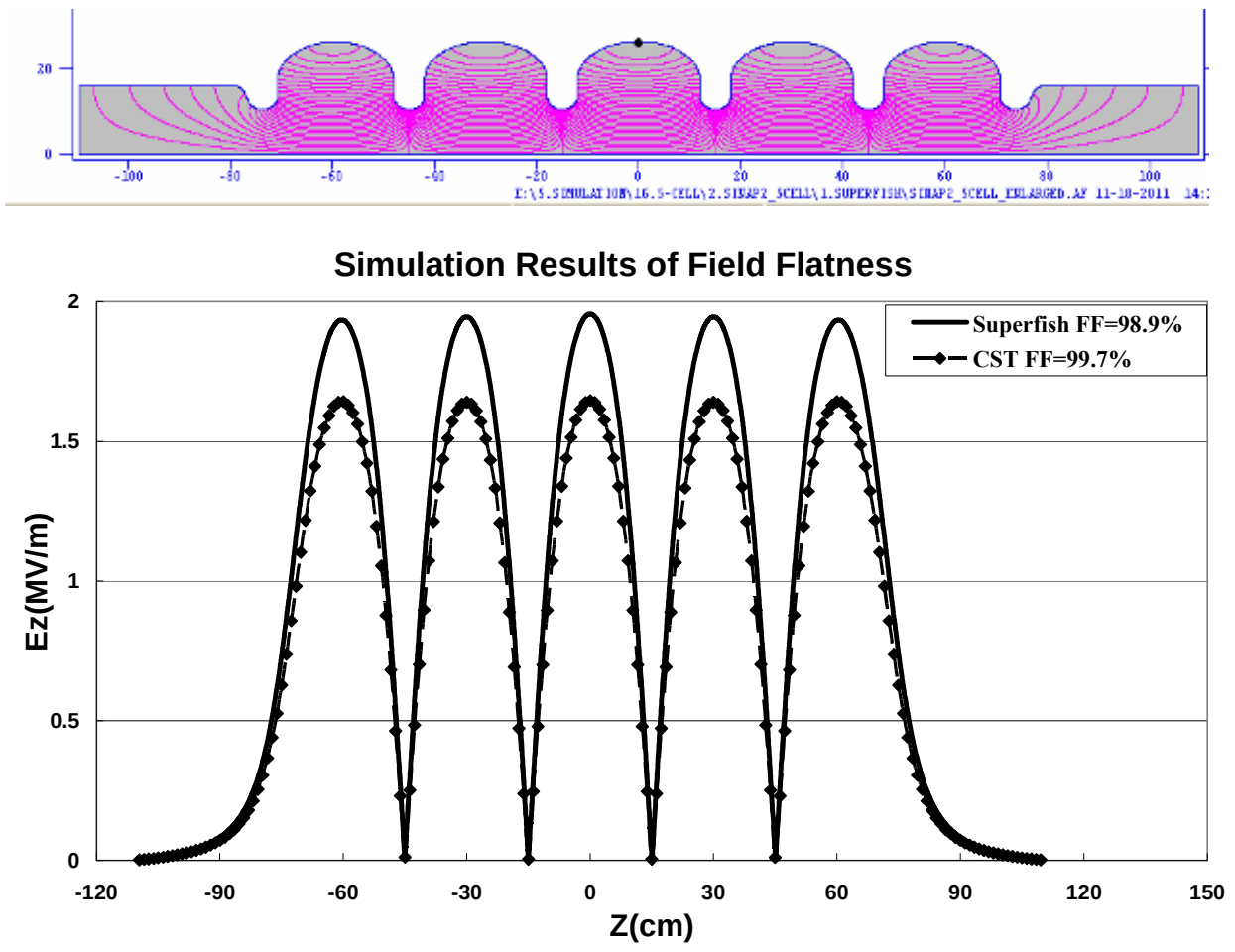

Figure 2 Field of 500 MHz 5-cell cavity accelerating mode and field flatness by Superfish and CST.

Damping of higher order modes is an important issue for the high current accelerator especially to those facilities operating in continues wave mode, thus many kinds of damping schemes have been proposed and performed [18-25] such as the fluted [21] or enlarged [22] beam pipe to propagate HOMs, waveguide coupler [20,23] and coaxial coupler [24-25] and so on. The enlarged beam pipe scheme has the advantage of a relatively simpler structure and can propagate almost all the dangerous HOMs out to the beamline type damper with optimization. The cut-off frequency of the beam pipe with a 105 mm radius is 837 MHz for the TE11 mode and 1742 MHz for the TM11 mode. The frequencies of the lowest two passbands of the five cell cavity HOMs are around 580 MHz and 670 MHz, which are below the cut-off frequency of the beam pipe and will be trapped in the cells. Therefore in order to propagate almost all the HOMs but not the accelerating mode, the beam pipe radius was enlarged to 160 mm to have cut-off frequency 549 MHz of TE11 mode. The longitudinal wake loss factor, which decides the average monopole mode HOM power in the cavity, was calculated by ABCI. Figure 3 shows the real part of the longitudinal impedance of monopole modes and the integrated loss factor which is 3.75 V/pC for 2 mm Gaussian bunch.

Multipacting is another key issue to be carefully handled when developing a superconducting cavity. The Fishpact code was used to simulate the whole cavity resulting in an enhanced counter function less than 0.2 with accelerating gradient up to 20 MV/m, as shown in figure 4, which means there was no multipacting area. It was proven partly by the later vertical test result that there was no multipacting at field up to 5 MV/m.

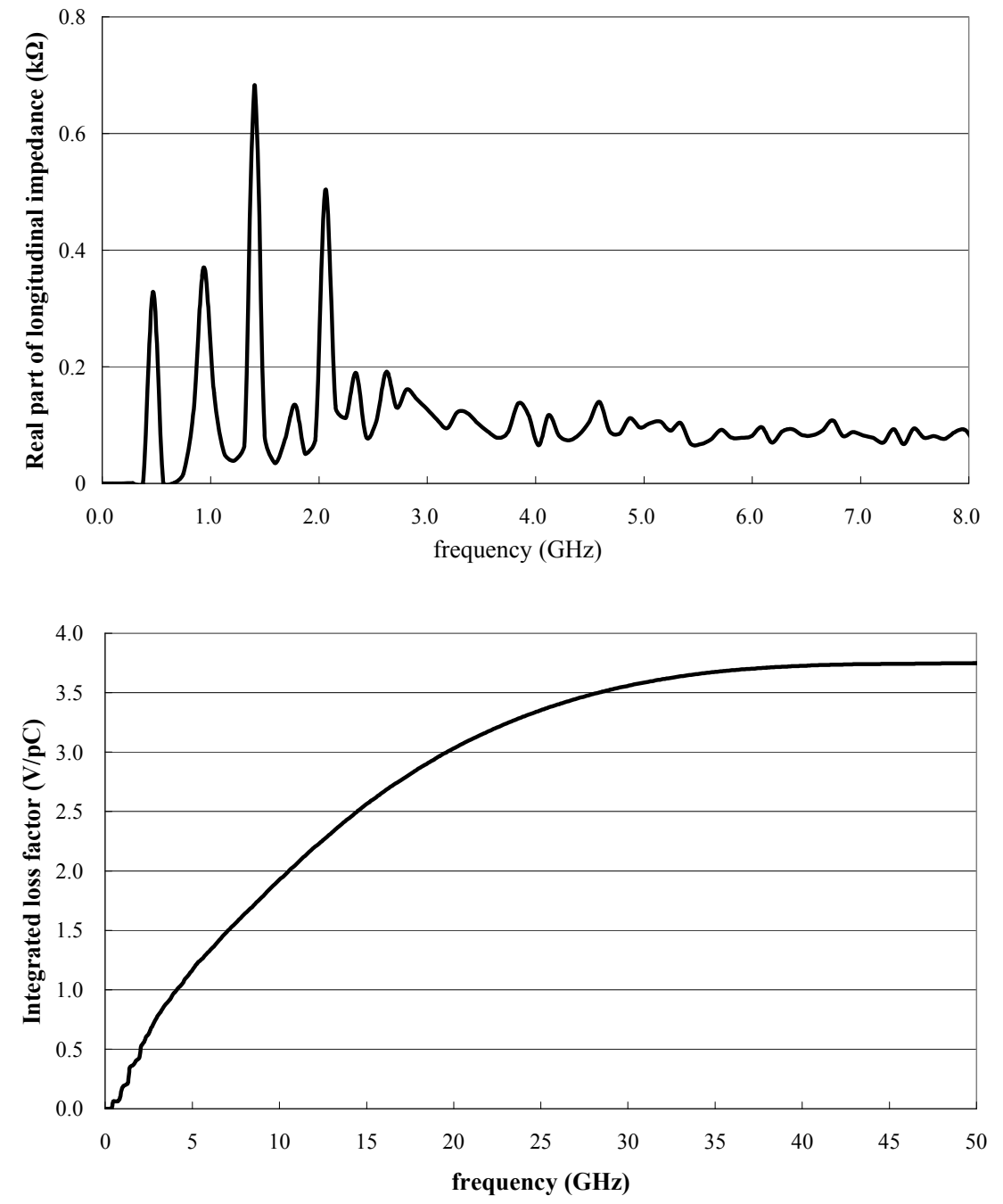

Figure 3 Real part of longitudinal impedance and integrated loss factor with frequency.

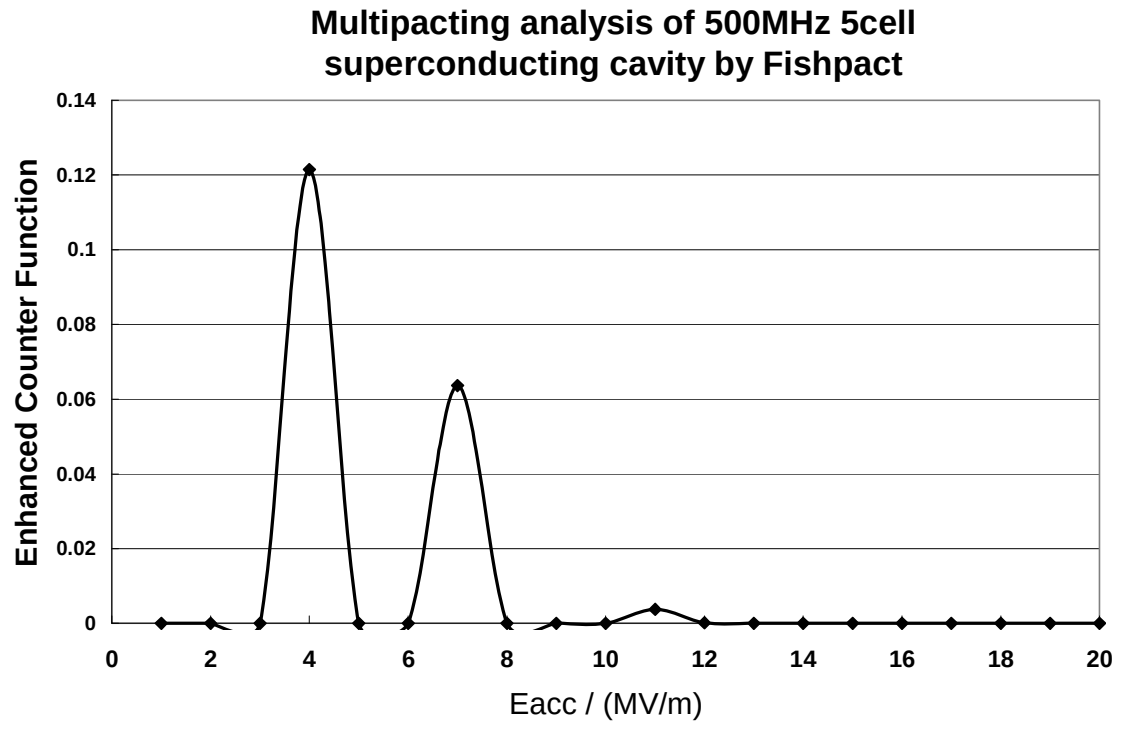

Figure 4 Enhanced counter function simulation results by Fishpact.

## 3 Fabrication and surface preparation

The deep-drawing together with the electron beam wielding (EBW) technology was adopted to fabricate the niobium half cells and the seamless beam pipe. This kind of method has been worldwide adopted in niobium cavities fabrication [24, 26]. One copper prototype cavity using the same method was built in order to verify the fabrication techniques and to explore the measurement method. Although the fabrication procedure was succeeded in deep-drawing the copper half cells, there happened cracks around the beam iris when pressing the niobium half cell. The selected fine grain pure niobium sheets with 3.2 mm in thickness and RRR (the residual resistance ratio) >300 had been annealed which turned out not be the reason. After conquering the cracks problem around the beam iris by separating the deep-drawing into two steps, the half cells were shaped successfully which could be seen in Figure 5. The trimming method was carried out to make the half cells dimensions to fit the frequency and field flatness requirements. The seamless beam pipe could be assembled to the NbTi flange easily and could avoid the EBW seam cross section around the flanges. However, it challenged the design complexity of the fabrication dies and the techniques because of its long length and the yield strength of the niobium.

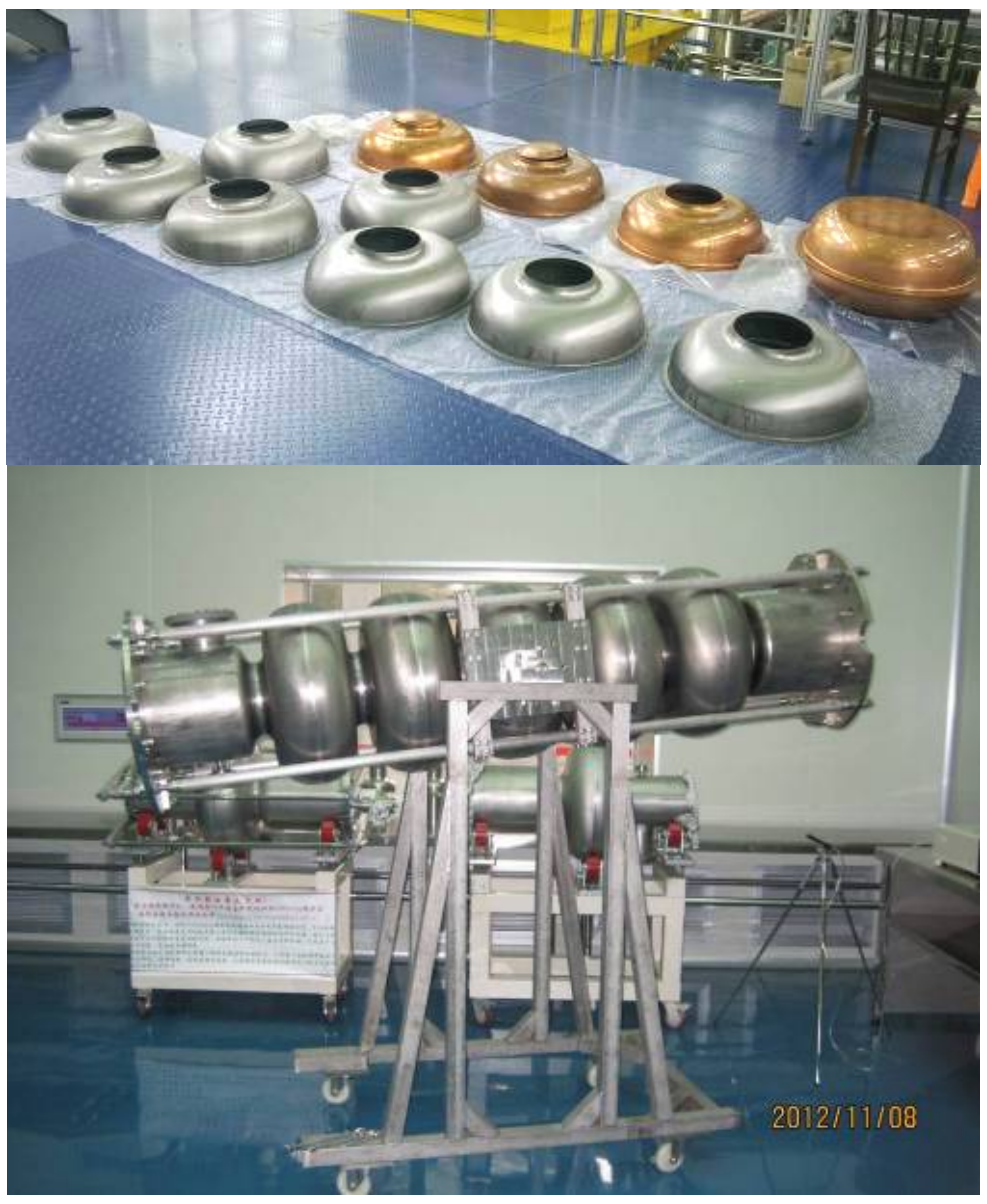

Figure 5 The formed half-cells and the successfully fabricated 500 MHz 5-cell niobium cavity.

The niobium half cells and beam pipes were cleaned, degreased carefully and rinsed with ultra-pure water before EBW to remove the oil and dust on the surface, especially on the welding joints. Then EBW was carried out with the vacuum pressure better than $5\times10^{-5}$ Torr. Figure 5 shows the formed half cells and beam pipes before EBW and the EBWed 500 MHz 5cell niobium prototype cavity.

With the successful fabrication of the cavity, different means of surface preparations were applied for smoothing the inner surface, which was included as follows:

- Measurement of the cavity dimensions
- Barrel polishing around 100 μm
- Heavy Buffered chemical polishing (BCP) around 100 μm with a ratio of $HNO_3$:HF:$H_3PO_4$ =1:1:2 by volume
- High pressure rinsing (HPR) with 18 MΩ-cm and 8 MPa ultra-pure water
- 680℃ annealing for 3 hours to remove hydrogen following heavy chemistry and distress
- Light BCP around 20 μm
- HPR again more than 48 hours because of the long length and large volume
- Clean room assembly
- Low temperature baking (~110 ℃) under vacuum better than 1E-8 mbar lasted more than 60 hours

## 4 Room temperature measurement

The cavity field flatness was adjusted [27] and the cavity rf performance at room temperature was measured after the annealing and before the light BCP step. The good field flatness will provide an equal kick in each cell by the π accelerating mode and will benefit the maximum accelerating voltage and minimum peak surface. The large cell-to-cell coupling and the low loss shape at the slope wall helped to reach field flatness better than 98% with the established pre-tuning equipment based on bead-pull method. Figure 6 shows the measurement setup and the final field flatness of the fundamental mode.

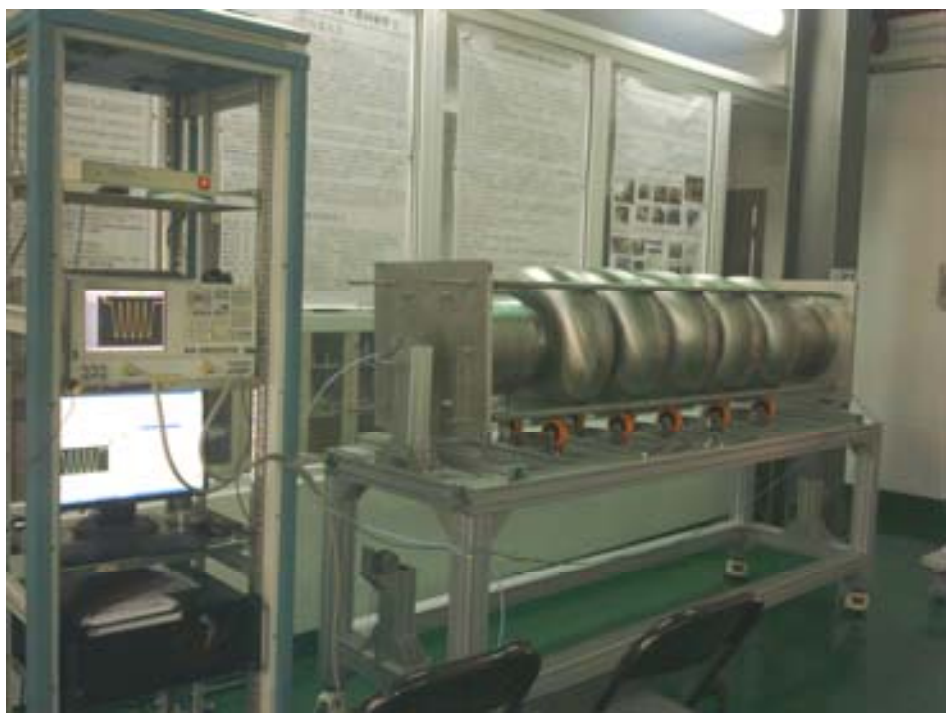

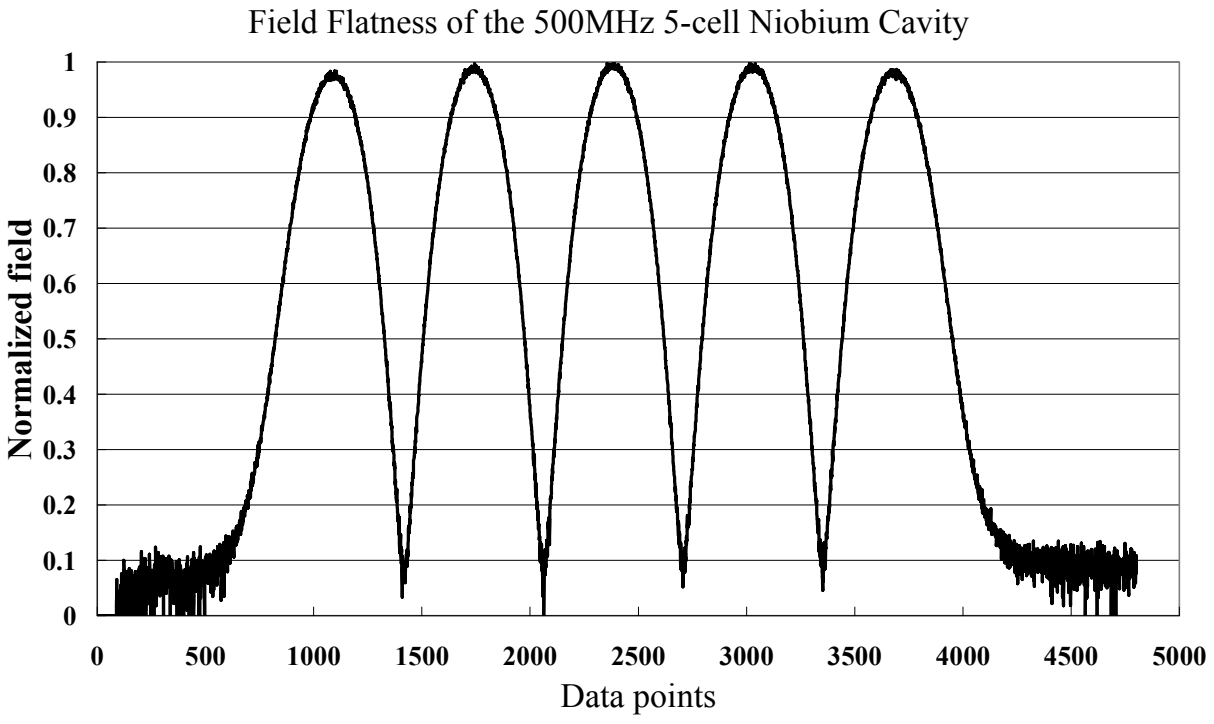

Figure 6 The pre-tuning frame for 500 MHz 5-cell niobium cavity and the final field flatness of fundamental mode.

After the cavity was tuned, the measurement result indicated that the field flatness would not change once the cavity length was fixed. The fundamental pass band of the niobium cavity was also measured to find the mode frequency separation between the accelerating π mode and the nearest 4π/5 mode was larger than 1.5 MHz which is large enough for the required cavity detuning in the operation.

## 5 Vertical test at 4.2K

The vertical test of 500 MHz 5-cell niobium cavity was carried out on the in-house designed and constructed facility [15] at SINAP. A new vertical test cryostat was fabricated with height up to 4.6 m for the 500 MHz 5-cell cavity. It is composed of helium vessel, outer vessel, thermal shielding vessel rounded with multi-layer super-insulation and liquid nitrogen cooling pipes. The magnetic shielding was inserted between the thermal shielding vessel and the outer vessel to reduce significantly the earth magnetic lower than 20 mGauss. The external magnetic should be also handled carefully because it can bring residual resistance to limit the cavity quality factor. The vacuum inside the cryostat during the test was obtained better than $1\times10^{-3}$ mbar. The static loss of the cryostat was around 10 W at 4.2K, which was good enough for vertical test.

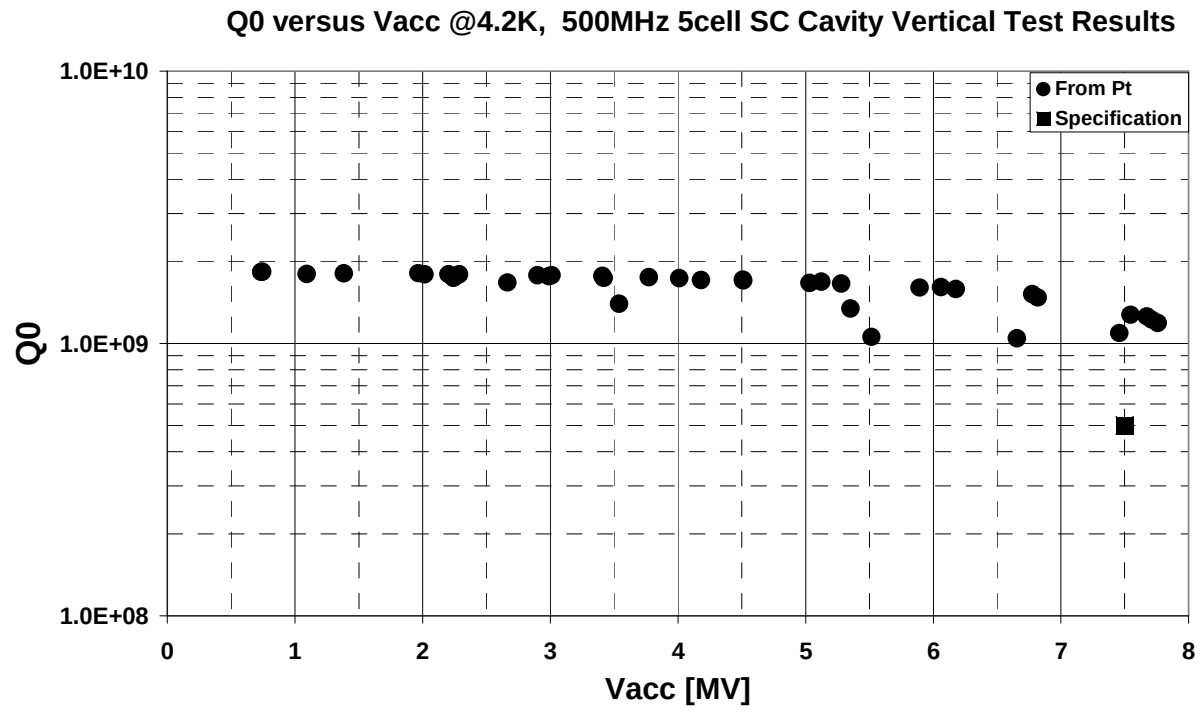

Figure 7 Vertical test results of the 500 MHz 5-cell superconducting cavity fabricated, surface prepared at SINAP. The cavity was fabricated by deep-drawing together with the electron beam wielding techniques. The inner surface was treated by mechanical polishing + 120 μm BCP + two hours 680℃ annealing + 20 μm slight BCP + HPR with ultra pure water + low temperature baking longer than 60 hours.

An adjustable input coupler was applied to vary the input coupling at a little bit greater than the critical point which is important especially for a cavity needs rf conditioning and good for obtaining the cavity performance. The vertical test at 4.2K has been carried out successfully and resulting in an accelerating voltage reached as high as 7.5 MV with unloaded quality factor better than $1\times10^9$. Two signal pickups $P_t$ and $P_{15D}$ with different coupling strength were equipped on the cavity during the test and both of them indicated the coincident results. There was no cold leak happened after the cavity emerged in the liquid helium which proved the electron beam wielding and the whole assembly sequences were correct. There was no obvious multipacting happened from the low field to the 7.5 MV voltage which meet the simulation results. Figure 7 shows the 4.2K vertical test result of our fabricated 500 MHz 5-cell cavity.

However, when a little more power was feed into the cavity, there arose a heavy thermal breakdown which limited the cavity accelerating voltage. It was suspected that somewhere on the inner surface or the wielding seam should have a foreigner to cause this thermal breakdown. The cavity will be disassembled in the future to find out where the impurity locates, and further surface treated and vertical tested again.

## 6 Conclusion & Discussion

The studies of a low loss type 500 MHz 5-cell superconducting niobium cavity including the design, fabrication, surface preparation and vertical test have been carried out. The 500 MHz 5-cell cavity has large beam apertures which can ensure lower HOM impedance and a promising high beam current threshold. The vertical test results indicate that the cavity accelerating voltage reaches as high as 7.5 MV with the quality factor better than $1\times10^9$ and there is no multipacting or field emission. The cavity accelerating voltage is limited by a quench.

## Acknowledgements

We would like to thank professors Robert Rimmer, Tom Powers from JLab, Wang Guangwei from IHEP for their discussion on the cavity forming and vertical test. We also thank professors ZHAO Zhentang and DAI Zhimin from SSRF for their persistent supports on developing the superconducting cavity.

## References

[1]Calaga R, Linear beam dynamics and ampere class superconducting RF cavities @ RHIC, Ph.D thesis, Stony Brook University, 2006.
[2]Rimmer R, Bundy R, Cheng G, et al., JLab high current CW cryomodules for ERL and FEL applications, WEPMS068, Proceedings of PAC07, Albuquerque, New Mexico, USA.
[3]Xu W, Ben-Zvi I, Calaga R, et al., High current cavity design at BNL, Nuclear Instruments and Methods in Physics Research A 622 (2010) 17-20.
[4]Umemori K, Furuya T, Sakai H, et al., Results of vertical tests for KEK-ERL 9-cell superconducting cavity, WEPEC030, Proceedings of IPAC10, Kyoto, Japan.
[5]Sekutowicz J, Kneisel P, Ciovati G, et al., Low loss cavity for the 12 GeV CEBAF upgrade, Jlab TN-02-023.
[6]Li Y M, Zhu F, Quan S W, et al., The design of a five-cell high-current superconducting cavity, CPC (HEP&NP), 2012, 36 (1):74-79.
[7]Wei Y L, Liu J F, Hou H T, et al., Design of large aperture 500MHz 5-cell superconducting cavity, Nuclear Science and Techniques 23 (2012) 257-260.
[8]Liu Z C, Gao J, Jin S, et al., High current superconducting cavity study and design, WEPWO016, proceedings of IPAC2013, Shanghai, China.
[9]Valles N, Liepe M, Furuta F, et al., The main linac for Cornell's energy linac : cavity design through horizontal cryomodule prototype test, Nuclear Instruments and Methods in Physics Research A, 734 (2014) 23-31.
[10]Carr G L, Martin M C, McKinney W R, High-power terahertz radiation from relativistic electrons, Nature, vol 420, 14 November 2002.
[11]Gruner S M, Bilderback D H, Energy recovery linacs as synchrotron light sources, Nuclear Instruments and Methods in Physics Research A 500 (2003), 25-32.
[12]Petenev Y, Atkinson T, Bondarenko A V, et al., Feasibility study of an ERL-based GeV-scale multi-turn light source, MOPPP016, Proceedings of IPAC2012, New Orleans, Lousiana, USA.
[13]Nakamura N, Review of ERL projects at KEK and around the world, TUXB02, Proceedings of IPAC2012, New Orleans, Lousiana, USA.
[14]Padamsee Hasan, Knobloch Jens, Hays Tom, RF superconductivity for accelerators, Cornell University, Ithaca, New York.
[15]Liu J F, Hou H T, Mao D Q, et al., Great progress in developing 500MHz single cell superconducting cavity in China, Science China, Physics, Mechanics & Astronomy, December 2011, Vol.54, Suppl. 2: s169-s173.
[16]http://laacg.lanl.gov/laacg/services/download_sf.phtml, Los Alamos National Laboratory poisson superfish codes.
[17]Microwave Studio manual, Computer simulation technology AG.
[18]Marhaugser F, HOM damping efficiency of various SRF coupler schemes, HOM damping workshop in SRF cavities, 2010.
[19]M.Liepe, J.Knobloch, Superconducting RF for energy recovery linacs, Nuclear Instruments and Methods in Physics Research A 557 (2006) 354-369.
[20]Rimmer R, Wang H, Wu G, Towards strongly HOM damped multi-cell RF cavities, proceedings of particle acceleration conference 2003.
[21]Padamsee H, Barnes P, Chen C, et al., Design challenges for high current storage rings, particle accelerators, 1992, Vol. 40, pp: 17-41.
[22]Mitsunobu S, Asano K, Furuya T, et al., Activities of superconducting cavity for KEK B-factory, proceedings of the sixth workshop on RF superconductivity.
[23]Rimmer R, Waveguide HOM damping studies at JLab, HOM damping workshop in SRF cavities, 2010.
[24]Aune B, Bandelmann R, Bloess D, et al., Superconducting TESLA cavities, physical review special topics – accelerators and beams, volume 3, 092001 (2000).
[25]Xu W C, HOM coupler design for high current SRF cavities, HOM damping workshop in SRF cavities, 2010.
[26]Saito K, Superconducting RF- basics for SRF technology, ILC 2nd summer school lecture.
[27]Tang Zheng-Bo, Ma Zhen-Yu, Hou Hong-Tao, et al., Frequency control and pre-tuning of a large aperture 500MHz 5-cell superconducting RF cavity, Nuclear science and techniques 25, 030102 (2014)